\documentclass[conference,compsoc]{IEEEtran}

\usepackage{eurosym}
\usepackage{verbatim}
\usepackage{subcaption}
\usepackage[normalem]{ulem}
\usepackage[hyphens]{url}
\usepackage{balance}
\newcommand{\urlBiBTeX}[1]{\url{#1}} 

\usepackage{xcolor}
\usepackage{colortbl}
\definecolor{Gray}{gray}{0.925}

\usepackage{tikz}
\usepackage{pgfplots}
\pgfplotsset{compat=1.10}

\usepackage[textwidth=1.5cm,textsize=tiny]{todonotes}
\newcommand{\TODO}[1]{\todo[inline]{#1}}

\usepackage{amsmath}
\usepackage{amssymb}

\usepackage{float}

\begin{document}

\setlength{\marginparwidth}{1.5cm}


\title{Economic Analyses of Security Investments on Cryptocurrency Exchanges}

\author{
\IEEEauthorblockN{Benjamin Johnson}
\IEEEauthorblockA{Technical University of Munich \\
Munich, Germany \\ 
benjamin.johnson@tum.de
}
\and
\IEEEauthorblockN{Aron Laszka}
\IEEEauthorblockA{University of Houston \\
Houston, TX, USA \\
alaszka@houston.edu}
\and
\IEEEauthorblockN{Jens Grossklags}
\IEEEauthorblockA{Technical University of Munich \\
Munich, Germany \\ 
jens.grossklags@in.tum.de}
\and
\IEEEauthorblockN{Tyler Moore}
\IEEEauthorblockA{The University of Tulsa \\
Tulsa, OK, USA \\ 
tyler-moore@utulsa.edu}
}


\maketitle

\begin{center}
   Published in the proceedings of the\\ 2018 IEEE International Conference on Blockchain.
   \end{center}

\begin{abstract}
Cryptocurrency exchanges are frequently targeted and compromised by cyber-attacks, which may lead to significant losses for the depositors and closure of the affected exchanges. These risks threaten the viability of the entire public blockchain ecosystem since exchanges serve as major gateways for participation in public blockchain technologies. 

In this paper, we develop an economic model to capture the short-term incentives of cryptocurrency exchanges with respect to making security investments and establishing transaction fees. Using the model, we derive conclusions regarding an exchange's optimal economic decisions, and illustrate key features of these conclusions using graphs based on real-world data.  Our security investment model exhibits horizontal scaling properties with respect to reducing exposure to losses, and may be of special interest to exchanges operating in markets with high price volatility.

\end{abstract}



\section{Introduction}
\label{sec:intro}

Public blockchain technologies and cryptocurrencies are inherently linked together.  The value of a public blockchain protocol gives value to those tokens which facilitate participation in the protocol.  Conversely, if the tokens required to use a blockchain protocol are difficult to monetize, the technology may lose value, or in extreme cases, may cease to function at all. This connection highlights the importance of cryptocurrency exchanges to the blockchain ecosystem.

Since the inception of Bitcoin in 2008, blockchain technologies and their associated cryptocurrencies have been on a spectacular (but volatile) path to success~\cite{nakamoto2008bitcoin}.  Bitcoin in particular, and blockchain technologies more generally, have fundamentally changed how users, businesses and financial intermediaries, as well as governments perceive the notion of virtual currencies \cite{ECB12}. 

Compared to traditional currencies issued by governments, virtual currencies are less regulated; their issuing technologies are generally controlled by programmers and math, rather than state agents; and their base of use and acceptance is still primarily a self-selected group of users and organizations. However, despite a bottom-up development approach, the practical constraints of sustaining  a cryptocurrency ecosystem have fostered the creation of a number of entities, such as mining pools, in which individual miners self-organize to smooth the financial risk of individualized mining operations. 

Similarly, a global and diverse ecosystem of cryptocurrency exchanges has grown into existence. On these third-party intermediaries, users can trade cryptocurrency for real money (or vice versa) or other cryptocurrencies \cite{Bhaskar15}. As such, exchanges facilitate inter-network transactions, but can also serve as a platform for financial speculation and arbitrage \cite{Glaser14,Kristoufek15}. In addition, exchanges serve as major gateways to the blockchain ecosystem. In fact, the purchase of any cryptocurrency coins on an exchange is often the first participatory step for consumers interested in joining the blockchain ecosystem.

At the time of writing, the top 5 exchanges cover over 50\% of the overall market share and trade a sizable number of cryptocurrencies; however, the entire cryptocurrency ecosystem consists of well over 50 exchanges with considerable variation in supported currencies and trading volume. Partly, this diversity can be explained by jurisdictional boundaries and the convenience to operate in a local currency. At the same time, market efficiency considerations would perhaps suggest an even higher degree of concentration than what is observed in practice. 

In particular, we observe that a higher level of concentration is under threat due to security issues. Because of their central role and currency-holding function, exchanges are a major target for insider fraud and external security compromises, which may lead to significant losses for the depositors and closure of the affected exchanges. 
For a three-year period ending in January 2013, Moore and Christin observed a 45\% closure rate for exchanges \cite{MooreC13fc}.  A follow-up study extending into 2015 calculated an overall exchange failure rate of 48\%~\cite{moore2018revisiting}. 

One of the most frequently cited examples of cryptocurrency exchange failures is the alleged case of fraudulent withdrawals from the Mt. Gox exchange, which led to its eventual bankruptcy and ongoing legal action. At its peak, Mt. Gox was facilitating over 70\% of all Bitcoin transactions. But Mt. Gox is not an isolated incident. 
More recently, in 2016, the Bitfinex exchange suffered a security breach that resulted in 119,756 bitcoins, which were worth around \$70 million at the time, being stolen from user accounts on the exchange~\cite{fortune2016bitcoin}.
Bitfinex eventually reimbursed the affected users, reinforcing their trust in the exchange, and it remains one of the largest exchanges in operation.

\textit{Such observations motivate our objective for this paper: to better understand the trade-offs between financial aspects of running a cryptocurrency exchange and potential investments in security.}
To address this goal, we investigate the market for cryptocurrency exchanges from an economic perspective. 
In particular, we develop an economic model for studying the day-to-day business incentives of an up-and-coming cryptocurrency exchange.

After motivating the tradeoffs of the model with respect to security investments and transaction fee rates, we conduct a decision analysis to determine an exchange's optimal choices in terms of a fixed set of parameters.
We then motivate specific values of those parameters using publicly available data involving cryptocurrency markets, and produce a set of illustrations for the optimal investments predicted by our analyses.

Finally, we discuss the recent market volatility in the price of various cryptocurrencies and its implications for the business operations of cryptocurrency exchanges. In particular, the optimal strategy for distributing loss exposure that arises from security investments within our model represents a form of horizontal scaling.  
We believe that implementing such a strategy could help cryptocurrency exchanges to make security investments at a rate that matches their fast-paced operating conditions, taking market price volatility into consideration.

%
%

Understanding incentive misalignments in the market of cryptocurrency exchanges is of critical importance for the current and future viability of public blockchain ecosystems. Consumers need to be able to rely on the availability and safety of their deposited assets, but at the same time a diverse system of exchanges can support goals, such as diversification and specialization.
We expect that our research contributes to a critical debate on optimal market structure and realistic security expectations.

\textbf{Roadmap:} In Section~\ref{sec:related}, we provide an overview of related work about cryptocurrency exchanges and economic analyses of cryptocurrency markets. In Section~\ref{sec:incentive}, we discuss the incentives that exchanges face in order to motivate our economic model, which is presented in Section~\ref{sec:model}. We conduct a decision analysis of our model in Section \ref{sec:analysis}, and 
present numerical results in Section~\ref{sec:numerical}. In Section \ref{sec:discussion}, we discuss the implications of distributed loss exposure strategies specific to cryptocurrency exchanges. Finally, we offer concluding remarks in Section~\ref{sec:concl}.


\section{Related Work}
\label{sec:related}

A growing number of comprehensive review and overview articles address Bitcoin, and cryptocurrencies in general \cite{Boehme14,Bonneau15}. Likewise, our work is relevant to the broader field of the economics of security (e.g., \cite{laszka2012survey}). In our discussion of related work, we focus on the functional, financial, and security aspects of cryptocurrency exchanges, and on economic models of virtual currencies.

\subsection{Research on Cryptocurrency Exchanges}


Most closely related to our work, Moore and Christin study factors for success and failure of Bitcoin exchanges with a three-year dataset ending in January 2013 \cite{MooreC13fc}. They discuss the reasons behind the observed 45\% closure rate, including security breaches. Statistical analysis suggests that more popular exchanges are attacked more often, but small exchanges are overall more likely to shut down. A more recent follow-up study provides a more detailed longitudinal analysis on newer data while confirming the same overall message~\cite{moore2018revisiting}.

Gandal et al. used data from the Mt. Gox Bitcoin currency exchange and determined how a single entity was able to manipulate prices in a highly significant fashion \cite{Gandal18}. Feder et al. also used Mt. Gox data to understand the impact of a DDoS attack on trade volume and the frequency of large trades \cite{Feder18}. 

A critical aspect of the viability of cryptocurrency exchanges is to limit the risk of bankruptcy or insider fraud. One mitigating factor is proof of solvency, which signals that deposits are correctly accounted for. This line of work includes Decker et al., who propose to use Trusted Platform Module (TPM), a secure cryptographic co-processor, to prove solvency~\cite{decker2015making}. Dagher et al.\ introduce Provisions, a privacy-preserving proof of solvency~\cite{dagher2015provisions}. This approach camouflages the total assets and liabilities associated with an exchange, while proving that assets are strictly greater than the accumulated liabilities.
Jain et al. study the optimization problem faced by exchanges when they have to decide how much of their holdings to keep in online storage to serve demands quickly, or to keep reserves offline (i.e., in comparatively safer cold storage) \cite{Jain15}.

Cryptocurrency exchanges have also been a fruitful ground for empirical finance studies. Gandal and Halaburda as well as Pieters and Vivanco examine exchange rates and arbitrage opportunities related to Bitcoin exchanges \cite{Gandal14,Pieters15}.
Glaser et al. investigate Bitcoin transaction and network volume as well as price formation with a particular focus on understanding intra-network transactions and on-exchange trading. Their analysis suggests that less informed (new) consumers are likely to hold on to bitcoins as an investment rather than using it for intra-network transaction purposes \cite{Glaser14}. Likewise, Kristoufek studies to which degree Bitcoin represents a speculative or conventional monetary asset \cite{Kristoufek15}. Bitcoin price formation has also been studied by Brandvold et al. \cite{Brandvold15}.
%


Despite the growing depth and diversity of research related to cryptocurrency exchanges, to the best of our knowledge, this paper is the first to
attempt modeling the incentives of cryptocurrency exchanges for security investments.

\subsection{Economic Models of Virtual Currencies}


Johnson et al. and Laszka et al. study the competitive (and at times adversarial interaction) of Bitcoin mining pools with different modeling approaches \cite{johnson2014game,Laszka15bitcoin}. They find that more sizable mining pools are more often victims of Distributed Denial of Service attacks, but are also more often the instigators. They also study the long-term impact of such attacks. Addressing a similar problem, Kroll et al. investigate the stability of mining on a public blockchain if an outsider has motivation to destroy the associated currency \cite{Kroll13}. More specifically, their ``Goldfinger'' attack compares on a high level the collective benefit of Bitcoin mining with some externally given incentive to destroy the economy altogether. 

Eyal and Sirer analyze to which degree Bitcoin may be manipulable by a colluding group of miners \cite{Eyal14}. Houy provides a complementary analysis showing that not only proof-of-work transaction validation, but also proof-of-stake validation is vulnerable \cite{Houy14}. Rather than assuming adversarial competition, Lewenberg et al. study how earnings of mining pools can be fairly distributed from a cooperative perspective and highlight problematic parameter constellations \cite{Lewenberg15}.


To the best of our knowledge, studies of the trade-off between financial aspects of running a cryptocurrency exchange and security investments are currently unavailable. We address this literature gap with our work.


\section{Incentives of Exchanges and Users}
\label{sec:incentive}

\subsection{Incentives of Exchanges}

Exchanges want to maximize profits, and at the very least, stay in business. To reach these goals, they need to achieve a certain level of transaction volume while maintaining sufficient operational security.

A cryptocurrency exchange is a two-sided market. Exchanges match sellers with buyers, and to be successful, an exchange should be attractive to both. For sellers, higher transaction volumes and liquidity are important, since they help ensure quick sales at fair market prices. Security is also paramount since the cryptocurrency is transferred to and stored on the exchange before selling to a buyer. For buyers, competitive transaction fees are crucial, but so is offering an easy way to transfer local currency into the cryptocurrency sold on the exchange. 
Given differing regulations across countries, some exchanges have specialized in serving certain markets. For example, at the time of this writing, the two leading exchanges converting between BTC and USD are Bitfinex and GDAX, 
both of which only accept USD as input. 
In addition to the difficulties of complying with local laws, such geographic dominance may also reflect the network effects of matching buyers and sellers of the same currency on particular platforms.

Like all two-sided markets, cryptocurrency exchanges exhibit positive externalities. This means that the value of an exchange grows as more users participate. In this case, having more users translates to higher transaction volumes, which makes the exchange more attractive to both buyers and sellers. Higher volume can also lead to higher revenue if the transaction fees are held constant, or lower transaction fees if the savings are passed on to customers.

Investing in security serves multiple purposes. Naturally, it protects against breaches, which are notoriously common amongst exchanges. Since there is usually very little recourse 
for exchanges after an attack, suffering a breach can deal a devastating blow. 
Prior work has established that transaction volume is positively correlated with experiencing a breach~\cite{MooreC13fc}, which makes sense because more successful exchanges present more valuable targets for attackers. Consequently, the need for security investment grows as an exchange becomes more popular. 

However, the aforementioned study also showed that low transaction volume can present an existential threat to the exchanges as well. When smaller exchanges are attacked, they are less likely to be able to recover, and may be forced to shut down. For example, the low-volume Polish exchange Bitmarket.eu operated for nearly 18 months, but closed following a breach and did not appear to reimburse customers. 

\subsection{Incentives of Users}

Security can also be used as a differentiating feature to attract customers. Many exchanges offer two-factor authentication features to customers, for example. Some, such as Kraken~\cite{krakenaudit}, publicize the fact that they have undergone security audits, and many advertise that they store large amounts of reserves in cold storage. Note that while it is easy for customers to verify the amount stored in cold wallets, the guarantee is not absolute: exchanges may mistakenly expose their cold wallets on vulnerable computers or even lose the private key associated with the wallet. However, the point of these actions may be less about improving the exchange's actual security and more about strengthening the {\em perceived} security of the exchange. Prospective customers are naturally concerned about the security of any deposits kept at exchanges, given multiple past examples where the customers of failed exchanges have lost money. Public displays of security investments can reassure customers and encourage would-be attackers to seek easier targets.

This discussion highlights the importance of customer interests and preferences to which successful exchanges must cater. In addition to security, several other factors influence customer choice among exchanges. One key factor is the transaction fee charged. Frequent traders are more likely to be sensitive to transaction fees. Exchanges make direct comparisons on fees difficult---some offer flat fees while others create a schedule that favors frequent traders or rewards loyal customers. For example, Bitstamp's fees start at 0.25\% for traders whose monthly volume is less than \$20,000, dropping to 0.10\% for volumes exceeding \$20M. These techniques are designed to encourage customer ``stickiness.''

Also important is the ease with which a given cryptocurrency can be purchased using external currencies. This can vary greatly by country, by cryptocurrency, and by exchange. While nearly all exchanges accept wire transfers, these are seen as inconvenient and expensive. A few exchanges, such as Coinbase and Circle in the United States, do accept credit cards, but the associated fees may discourage casual users. Regulatory actions sometimes get in the way, however. For example, following its hack, US-based correspondent banks began blocking international USD wire transfers to Bitfinex~\cite{bitfinexusd}.





\section{Economic Model}
\label{sec:model}

\subsection{Overview}

In the previous section, we discussed a range of incentives that
influence the decisions of users in selecting exchanges, as well as
the options for differentiation among exchanges. We now translate
these high-level incentives into a simplified analytical model, designed to capture the day-to-day business incentives of a cryptocurrency exchange.
Our model focuses on two decisions made by the operators of an exchange: the investment in security measures to reduce its exposure to loss, and the choice of transaction fees on trades to generate revenue.
Table~\ref{tab:symbols} lists the symbols used in our model.

\begin{table}[h]
\caption{List of Symbols}
\label{tab:symbols}
\centering
\renewcommand*{\arraystretch}{1.15}
\setlength{\tabcolsep}{3pt}
\begin{tabular}{| c | p{7.15cm} |}
\hline
Symbol & Description \\
\hline
\multicolumn{2}{|c|}{Functions and Parameters} \\
\hline
\rowcolor{Gray} $P(t)$ & profit on day $t$ \\
$R(t)$ &  revenue on day $t$ \\
\rowcolor{Gray} $C(t)$ & costs on day $t$ \\
$C^B(t)$ & business expenses on day $t$ \\
\rowcolor{Gray} $C^R(t)$ & business risk on day $t$ \\
$V(t)$ & total volume of trades occurring within the exchange on day $t$ \\
\rowcolor{Gray} $p(t)$ & probability that an incident happens on day $t$ \\
$L(t)$ & loss that will occur in case an incident happens on day $t$ \\
\rowcolor{Gray} $\alpha_M$ & attack-probability constant \\
$\beta_M$ & risk-exposure constant \\
\rowcolor{Gray} $\gamma_{BS}$ & fraction of users who are willing to leave an established exchange for a lower transaction fee \\
\hline
\multicolumn{2}{|c|}{Choice Variables} \\
\hline
\rowcolor{Gray} $f$ & fee rate \\
$I(t)$ & security investment \\
\hline
\end{tabular}
\end{table}

\subsection{Profit}
A cryptocurrency exchange is an online business that allows its users to make two-party trades involving two different types of currencies, at least one of which is a cryptocurrency. 
As a business, their day-to-day operational decisions revolve around the notion of profit.  We define the daily profit $P(t)$ of an exchange to be the difference between its daily revenue $R(t)$ and its daily costs $C(t)$.
\begin{equation}
P(t)=R(t)-C(t) .
\end{equation}

\subsubsection{Revenue}
To keep the revenue model simple but realistic, we focus on the most common and typically largest source of revenue for exchanges, which is a transaction fee on trades, assessed as a percentage of the volume of the trade. For our modeling purposes, suppose that fee rate $f$ is some constant percentage rate\footnote{In practice, the fee rate often depends on the amount of currency being offered or asked, the type of transaction (e.g., limit order or market order), etc.
Our model abstracts away from the details of fee structures, and focuses on fee rates for average users in comparison with other exchanges.}
 (for example, 20 basis points or 0.2\%), and let $V(t)$ denote the total volume of all trades within the exchange occurring during day $t$ (expressed in the resolving currency of the exchange, e.g., USD).  Then, the exchange's daily business revenue $R(t)$ is given by 
\begin{equation}
R(t)=f\cdot V(t) .
\end{equation}

\subsubsection{Costs}
In our model, we divide costs into business expenses $C^B$ and business risks $C^R$. Business risks represent losses that might occur due to incidents. The daily costs may thus be expressed simply as
\begin{equation}
C(t)=C^B(t) + C^R(t) .
\end{equation}

\subsubsection{Security Investments}
An exchange incurs a variety of business expenses, all of which may be important if we consider the long-term strategies of an exchange to maintain liquidity and stay in business.  However, our approach here only considers short-term operational decisions involving security, which may be considered as just a subset of an exchange's real decision-making requirements.
In particular, our decision model is focused entirely on security and fees, and so the present cost model 
considers only the cost $I(t)$ of investing in security.
\begin{equation}
C^B(t) = I(t) .
\end{equation}

The purpose of security investments is to reduce the risk or the exposure to potential loss from incidents, such as security breaches and insider attacks.

\subsubsection{Business Risks}
Operating an exchange is a risky business. For example, hackers may compromise parts of the exchange software that allows them to steal user funds and/or business liquidity.  Rogue employees may also steal funds to similar effect.  
In our model, the risk-related cost for day $t$ is a product of two random variables $1_{p(t)}\cdot L(t),$ where $p(t)$ is the probability that an incident happens on day $t$, $1_{p(t)}$ is a Bernoulli random variable indicating whether or not the incident actually happens on day $t$, and $L(t)$ is the loss that will occur in case an incident happens.
Hence,
\begin{equation}
C^R(t) = 1_{p(t)}\cdot L(t) .
\end{equation}


\subsubsection{Summary}
Putting everything together, the profit of an exchange on day $t$ may be expressed as 

\begin{equation}
\label{eq:profit}
P(t)= f\cdot V(t)-I(t)-1_{p(t)}\cdot L(t) ,
\end{equation}
where the choice variables for an exchange are the fee rate $f$ and the daily security investment $I(t)$.

In the following subsections, we provide additional structure to the model, by constraining the relationships between security investment and expected outcomes, and between fee levels and volume.  Because our analysis focuses on day-to-day incentives for exchanges, we will sometimes omit the time variable $t$.

\subsection{Risk Valuation}
The daily risks of an exchange may be considered along two dimensions: probabilities and exposure.  What is the likelihood that an exchange gets breached on a particular day? And what is the potential loss if it does get breached on a given day? Our model will consider these concerns separately.

\subsubsection{Probability of Attack}
The economic approach to predicting events such as being hacked on the Internet, whose probabilities are difficult to measure, is based on the assumption that attackers are incentivized to attack the most valuable targets. A complementary extension of that principle, which we apply to our model, says that the probability of an exchange being breached is 
proportional to the volume of the exchange.  

There is practical evidence suggesting that this is indeed true. Research on exchange closures has shown a strong positive correlation between the volume of an exchange and whether that exchange experiences a security incident~\cite{MooreC13fc}. 

Our model will assume that the probability of an exchange being attacked is proportional to the relative volume of the exchange within the given market.  An exchange having a larger share of the market will receive more attention and face greater risks.
Let $M$ be an exchange market (i.e., a pair of currencies to be exchanged); let $V_M$ denote the total volume of this market, and let $V$ denote the volume of a specific exchange participating in $M$.  Our assumption says that 
$$p\sim\frac{V}{V_M}.$$

Finally, we let $\alpha_M$ denote the constant of proportionality.
The constant $\alpha_M$ depends on the market $M$, and we can estimate it by considering historical breach data (see Section~\ref{sec:parameters}). 
In sum, our constraining assumption gives
\begin{equation}
\label{eq:p}
p  = \frac{\alpha_M\cdot V}{V_M} .
\end{equation}

\subsubsection{Exposure to Loss}

The minimum exposure of an exchange throughout a given day is the amount of currency that is necessary to fulfill the trades on that day.  Even though it is possible to move assets to more secure storage locations (e.g., cold storage), some of the funds need to be accessible through the Internet for liquidity, and thus be exposed to the highest levels of risk.  In the best case, the loss exposure of an exchange on a given day is proportional to its daily volume.  However, it may be much worse.  If the exchange has poor security or exposes more of its assets than necessary, a breach could have a more catastrophic impact.  

Suppose that some amount of security investment $I^*$ is sufficient for reducing the loss exposure to exactly the daily volume of the exchange.  Our model treats $I$ as the entire costs of operating the exchange, so let us consider a somewhat hypothetical question.  How much could the exchange reduce its loss exposure by doubling its investment to $2 \cdot I^*$? 
By approximately doubling its entire operating cost, it could funnel half of its volume through a completely independent exchange system.
Then, each of these two exchange systems would expose exactly half the daily volume compared to before.  
The same logic applies to tripling the investment, and so on.
Consequently, we can assume that the loss exposure of an exchange is proportional to its daily volume divided by its security investment.  That is, 
%
%
%
$$L\sim \frac{V}{I} .$$

We let $\beta_M$ denote the constant of proportionality. This constant also depends on the market $M$, and we will estimate it by considering the price of white-label exchange software (see Section~\ref{sec:parameters}).
In sum,
\begin{equation}
\label{eq:L}
L=\frac{\beta_M\cdot V}{I} .
\end{equation}

Note that this security investment mechanism for distributing loss exposure exhibits the properties of horizontal scaling, similar to how additional web servers may be added to handle traffic congestion.

\subsection{Fees and Volume}

There is an inverse relationship between transaction fees and transaction volume, at least according to classical economic theory, which 
says that if the price of a product or service rises, fewer people will purchase it. 
Unfortunately, beyond that basic principle it is difficult to imagine the smooth equilibrium results of prices in competition to hold in regimes in which the economic fundamentals are as volatile as the exchange prices of today's cryptocurrencies.

Therefore, we are forced to take a somewhat less calculus-oriented approach to considering how, say, an incoming (initially small) cryptocurrency exchange $S$ would choose a value for its transaction fee $f_S$.  
Suppose that for a given market $M$, there is a large dominant exchange $B$, which charges a transaction fee level $f_B$.  Since $B$ is well-funded, feature-rich, and heavily marketed, it seems unlikely for a new exchange to attract any substantial market share whatsoever if they start with a fee level higher than that of $B$.  So the choices for $S$ can reasonably be restricted to lie in the interval $[0,f_B]$.  

Now consider also that in today's market, sophisticated users are likely to be more concerned about price volatility than they are about transaction fees.  After all, if we look at the daily statistics from a variety of cryptocurrency exchanges, we see daily variations in the prices reaching up to 10\% of the total price of the currency being traded.  This variation is almost two orders of magnitude greater than the typical transaction fees that are assessed on trades. So the risk of participating in a low volume exchange would seem to dominate the small variation in transaction fee levels between exchanges.

Therefore, we may safely assume that there is only a  small fraction of the users (say $\gamma_{BS}$) that are willing to leave an established exchange $B$ for a smaller exchange $S$ only on the basis of having a lower transaction fee.  For the sake of concreteness, let us assume that this small set of users are distributed uniformly in their sensitivity to fees, so that the volume of $S$ increases linearly from 0 to $\gamma_{BS}V_B$ as $f_S$ decreases from $f_B$ to zero. This assumption yields
\begin{equation}
\label{eq:VS}
V_S=\frac{\gamma_{BS} \cdot V_B\cdot(f_B-f_S)}{f_B} .
\end{equation}

By using this equation, we may calculate the anticipated volume $V_S$ of a new exchange in terms of its own fee level $f_S$, the fee level of the big exchange $f_B$, the volume of the big exchange $V_B$, and the parameter $\gamma_{BS}$.


\section{Decision Analysis}
\label{sec:analysis}

\subsection{Overview}
In this section, we analyze the model, presented in the previous section, to calculate the security investments and fee levels that maximize an exchange's daily profit as a function of the environmental parameters. 

Recall that the daily profit of an exchange was given in Equation \ref{eq:profit} by 
\begin{equation}
P(t)= f\cdot V(t)-I(t)-1_{p(t)}\cdot L(t) .
\end{equation}

Because we will focus on the day-to-day optimization, we can omit the variable $t$ and avoid the nondeterministic nature of the formula by evaluating the expected daily profit $E[P]$. To do this, we replace the random variable $1_{p}\cdot L$ with its expected value $pL$.

Putting it all together, our goal is to maximize the expected daily profit of the exchange, which is
\begin{equation}
\label{eq:EP}
E[P] = f\cdot V - I - p \cdot L .
\end{equation}




\subsection{Optimizing Security Investments}
\label{sec:securityAnalysis}

We can optimize the investment levels for expected profit by expanding the terms of Equation \ref{eq:EP} and using calculus. 
Expanding those terms we have
\begin{align*}
E[P] &= f \cdot V - I - p \cdot L \\
&=f\cdot V - I - \frac{\alpha_M \cdot V}{V_M}\cdot \frac{\beta_M \cdot V}{I}.
\end{align*}
To find the optimum investment level, calculate the first derivative of profit with respect to security investment and set it equal to zero.
\begin{align*}
\frac{\partial E[P]}{\partial I}&=-1 + \frac{\alpha_M \cdot \beta_M \cdot V^2}{V_M\cdot I^2} = 0\\
I^2&=\frac{\alpha_M \cdot \beta_M  \cdot V^2}{V_M}\\
I&=V\sqrt\frac{\alpha_M \cdot \beta_M}{V_M} .
\end{align*}

We see that the optimal investment in distributed security is a 
proportion of the exchange's daily volume.  The proportion itself depends on the volume in the entire market, as well as the parameters describing the market's overall riskiness, and the cost of distributing exposure to losses. 

By adopting this level of investment, the expected profit of an exchange becomes 
\begin{align*}
E[P]&=f\cdot V - I - \frac{\alpha_M \cdot V}{V_M}\cdot \frac{\beta_M \cdot V}{I}\\
&=f\cdot V - V\sqrt\frac{\alpha_M \cdot \beta_M}{V_M}- V\sqrt\frac{\alpha_M \cdot \beta_M}{V_M}\\
&=V\left(f - 2\sqrt\frac{\alpha_M \cdot \beta_M}{V_M}\right) .
\end{align*}

%

\subsection{Optimizing Fee Levels}
In order to maximize revenue from the fee, an exchange $S$ should choose its fee level $f_S$ to maximize the portion of profit which it affects.  We will compute this in two ways, both of which may be well-motivated in some circumstances.

First, we compute the optimal fee level by ignoring the effects of security investment.  The rationale for this approach is that, when using real-world values for the parameters $\alpha_M$, $\beta_M$ and $V_M$, the magnitude of the side effect is very small compared to the typical market value for transaction fees, 
hence the effect of considering security is only noise on the approximate optimal value of the fee level.

To implement this approach, we simply approximate $\frac{\partial E[P]}{\partial f}$ by $\frac{\partial fV}{\partial f}$, substitute for $V$ using Equation \ref{eq:VS}, and set the result equal to zero. Doing this, we obtain
\begin{align*}
\frac{\partial (f_SV_S)}{\partial f_S}&= \frac{\partial}{\partial f_S}\left(f_S\cdot\frac{\gamma_{BS} \cdot V_B\cdot(f_B-f_S)}{f_B}\right)\\
0&=\frac{\gamma_{BS} \cdot V_B\cdot(f_B-2\cdot f_S)}{f_B}\\
f_S&=\frac{f_B}{2} .
\end{align*}
We see that in this case, the exchange's optimal fee rate is exactly half the rate set by the bigger exchange. 

Secondly, we compute the optimal fee value while considering the side effects of security investment.  This will help us to see how the dynamic might evolve for exchange markets in which either there is abnormally high risk, or in which the cost of operations is large compared to the total volume of the market.
To implement this approach, we put all of our equations together, and follow the strategy described above.
\begin{align*}
E[P]&=V_S\left(f_S - 2\sqrt\frac{\alpha_M \cdot \beta_M}{V_M}\right)\\
&=\frac{\gamma_{BS} \cdot V_B\cdot(f_B- f_S)}{f_B}\left(f_S - 2\sqrt\frac{\alpha_M \cdot \beta_M}{V_M}\right)\\
\frac{\partial E[P]}{\partial f_S}&=\frac{\gamma_{BS}V_B}{f_B}\left(f_B+2\sqrt\frac{\alpha_M \cdot \beta_M}{V_M}-2f_S\right)=0\\
f_S&=\frac{f_B}{2}+\sqrt\frac{\alpha_M \cdot \beta_M}{V_M} .
\end{align*}
Again in this case, we see that the optimal fee rate for exchange $S$ is similar to half the big exchange's rate, but here it should be slightly higher to account for the additional risk that comes with increased revenue. This additional risk essentially creates a drag on the effectiveness of the fee rate with respect to profit generation, so that the optimum profit level is not reached until there is a slightly higher fee rate.

\section{Numerical Analysis}
\label{sec:numerical}

\subsection{Overview}
In this section, we give values to the various parameters by using real world data involving market volumes, the relative sizes of big and small exchanges, and the historical probabilities of exchanges being breached. We interpret the results in terms of these figures, and provide numerical illustrations for the conclusions of our analyses.

\subsection{Parameter Estimations}
\label{sec:parameters}
The values for the parameters involving total market volume $V_M$, volume of the big exchange $V_B$, and fee level of the big exchange $f_B$ were estimated based on publicly-available data on exchange volumes for the Bitcoin economy~\cite{blockchaininfocharts}, 
using 
BitFinex as an example of a big exchange.  These estimates are $V_M=\$600$M, $V_B=\$200$M, and $f_B=0.002$, respectively.  

The fraction of users willing to leave a big exchange~$B$ for a smaller exchange $S$ based on fee levels, $\gamma_{BS}$, was estimated at 0.01 based on the fact that a median Bitcoin exchange by volume exchanges somewhat less than 1\% of the total market volume.


The parameter $\alpha_M$ measures the probability of a successful attack against one of the exchanges in a market on a given day.  Based on data for Bitcoin exchange attacks, from the aforementioned study by Moore and Christin \cite{MooreC13fc}, which reported 27 different exchanges that experienced a breach during a period of some 1918 days.  This event space yields an average daily probability of breach of roughly 1.4\%.

Lastly, the parameter $\beta_M$ is the unit cost of operating an additional risk-independent cryptocurrency exchange. We estimated this parameter by aggregating publicly-available pricing information for various managed solutions providing ``white label'' exchange software. For example, Draglet~\cite{dragletweb} 
offers a fully managed exchange software for around US \$75K per year.  Meanwhile, BTC Trader~\cite{btctraderweb} 
offers what seems to be a similar product requiring an initial license fee of US \$25K and 30\% of the transaction fees, with a minimum fee of US \$10K per month.  Averaging the (minimum) costs of these products, amortizing the flat fees over the course of one year, and expressing the result in units of dollars per day, yields an estimate for $\beta_M$ of approximately~\$300.

\subsection{Security Investment and Risks}

\begin{figure}[H]
\centering
\includegraphics[width=\columnwidth]{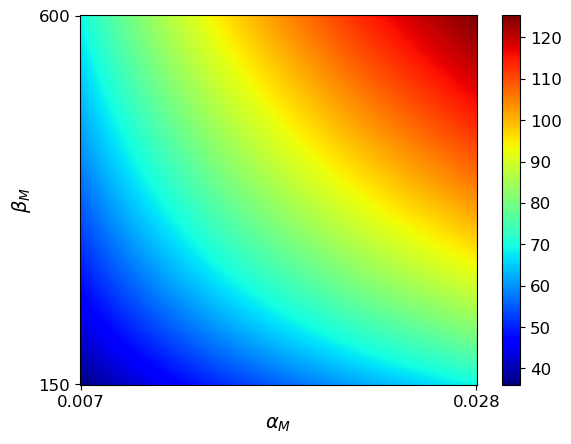}
\caption{Optimal security investment $I_S$ as a function of attack-probability constant $\alpha_M$ and risk-exposure constant~$\beta_M$. Here, $V_M=600$M, $V_B=200$M, $f_B=0.002$, $\gamma_{BS}=0.01$, and $f_S$ is the optimal fee rate.}
\label{fig:alphaM_betaM_opt_IS}
\end{figure}

Figure~\ref{fig:alphaM_betaM_opt_IS} shows the optimal security investment $I_S$ as a function of the attack-probability constant $\alpha_M$ and the risk-exposure constant~$\beta_M$.  The fee level was also optimized based on the optimal security investment, to be $f_B / 2 + \sqrt{\alpha_M \cdot \beta_M / V_M}$.
As expected, both higher attack probabilities and potential losses incentivize the exchange to invest more in security in order to mitigate the higher risk.
However, if either one of these factors is very low, then security risks are manageable even without significant investments. 

\subsection{Daily Profit and Choices}

\begin{figure}
\centering
\includegraphics[width=\columnwidth]{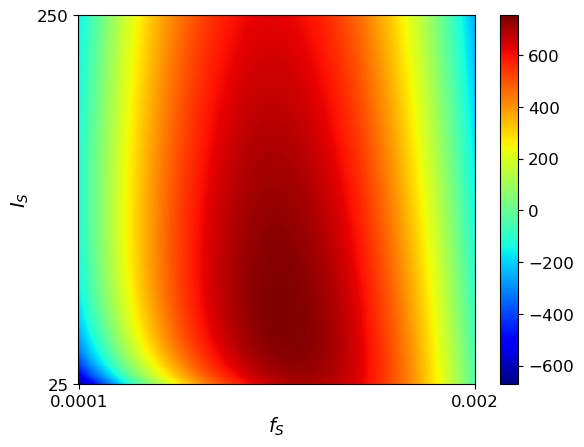}
\caption{Profit $P_S$ as a function of fee rate $f_S$ and security investment $I_S$.  Parameter values are 
$V_M=600$M, $V_B=200$M, $f_B=0.002$, $\gamma_{BS}=0.01$, $\alpha_M=0.014$ and $\beta_M=300$. 
}
\label{fig:fS_IS_PS}
\end{figure}

Figure~\ref{fig:fS_IS_PS} shows the exchange's daily profit as a function of its fee rate $f_S$ and security investment $I_S$.
The optimal fee rate and security investment are $f_S = 0.108\%$ and $I_S = \$69.0$.
Either increasing or decreasing the fee from this optimal level will gradually decrease the exchange's profit, either due to becoming not competitive in the market or due to being insufficient to collect significant revenue.
Decreasing security investments results in a rapid decrease in profitability due to the rapid increase in security risks.
However, increasing security investments decreases profitability slowly due to the slowly increasing business expenses.
This suggests that rational exchanges should not shirk security investments when they are uncertain about the optimal level.

%
%
%
%
%
%

\subsection{Security Investment and the Market}

\begin{figure}
\centering
\includegraphics[width=\columnwidth]{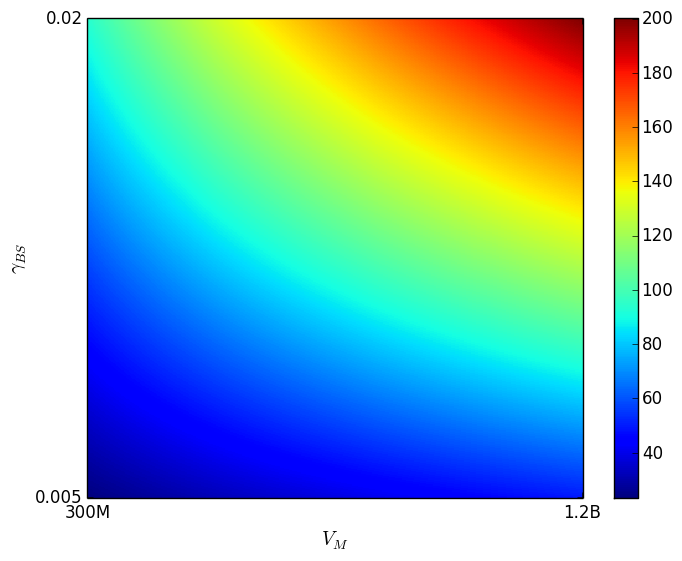}
\caption{Optimal security investment $I_S$ as a function of total market size $V_M$ and fraction of fee-sensitive users $\gamma_{BS}$. Here, $V_B=200$M, $f_B=0.002$, $\alpha_M = 0.014$, $\beta_M = 300$, and $f_S$ is the optimal fee rate.}
\label{fig:VM_gamma_opt_IS}
\end{figure}

Figure~\ref{fig:VM_gamma_opt_IS} shows the optimal security investment $I_S$ as a function of the total market size $V_M$ and the fraction of fee-sensitive users $\gamma_{BS}$.
For this plot, we assume that the fee level is always the optimal value $f_B / 2 + \sqrt{\alpha_M \cdot \beta_M / V_M}$.
As expected, larger markets lead to greater security investments due to the higher stakes in terms of risks.
We also observe that, interestingly, users being more sensitive to fee levels also results in greater security investments.

\section{Volatility and Implications}
\label{sec:discussion}

%
One of the most interesting features of the cryptocurrency exchange business is the phenomenon of huge price volatility.  This effect makes the economic analysis of cryptocurrency exchanges substantively different from traditional businesses, including traditional currency exchanges.  The high volatility essentially has the effect of placing a large proportion of the business revenue in the hands of the changing market.  

The extent to which exchanges may adjust their fee levels is already limited by their placement within an existing market.  
But even more generally than that, the factors that influence fee selection even in more stable exchange markets do not ordinarily have much to do with the specific valuations of the traded currencies.  
That is to say, the choice of fee level to maximize profit is largely independent of the currency valuation. 

However, the same cannot be said for the value of making security investments.  The effect on security is absolutely tied to the value of the assets being exchanged.  The pressure is in two similarly well-motivated dimensions.  First, in the risk dimension, the attention from attackers to the exchange market is proportional to the value of that market, because attackers tend to expend the most effort where the money is.  Similarly, in the loss dimension, the exposure of an exchange is inextricably linked to the market value of its traded assets. For example, consider an exchange behaving in exactly the same way under two environmental conditions, one in which the value of a traded asset experiences a five-fold increase in its market value.  In the high-value case, the loss exposure of the exchange is five times higher; and this happens without any change of actions by the exchange with regard to its security or business operation.


Given that security incentives for cryptocurrency exchanges \emph{should} increase roughly in proportion to the value of the exchanged currencies, it is reasonable to question whether our security technologies are equipped to scale at the pace at which many cryptocurrency exchange markets are moving.  Cryptocurrency exchanges over the past year have experienced business environments in which both their primary sources of revenue as well as their liabilities for potential loss of assets have increased and/or decreased more than ten-fold in the span of much less than a year. This paper presents a model of distributed security in which horizontal scaling of distributed exchange architectures can be used to reduce the exposure of traded volume to would-be hackers.  We are not aware of an existing technology that does this in real time; but our results suggest that such a technology would be useful to cryptocurrency exchanges to optimize a portion of their investment strategies. 





\section{Conclusions}
\label{sec:concl}

Cryptocurrency exchanges serve as the major gatekeepers for participation in public blockchain technologies, and are an important factor for the continued success and future growth of such technologies. 
Just as is the case for exchange markets for traditional currencies, it is imperative that cryptocurrency market participants can have trust in the reliability, liquidity, and security of their deposits. Previous research has shown that these criteria are not always matched as evidenced by numerous incidents of cryptocurrency exchanges closing their doors \cite{MooreC13fc}. Very recent examples of exchange closures (for a variety of reasons) include Coin.mx and Yacuna, while other exchanges such as Cryptsy are struggling to stay open. Some exchanges including Bitfinex have also weathered recent security attacks and managed to continue business operations with limited long-term impact for their customers.

Given the importance of cryptocurrency exchanges, it is prudent to better understand the fundamentals surrounding exchange risk management. For this purpose, we provided a discussion of incentives which impact the operations of an exchange. The observations guided our development of an economic model to study the trade-off between financial aspects of running a cryptocurrency exchange and security investments.

After presenting our model for investigating the daily profit of a cryptocurrency exchange, we applied this model to study the day-to-day business incentives of an exchange, with a specific focus on security investments and the selection of transaction fees. Our method of assessing the effect of security investments in terms of distributing loss exposure appears highly relevant in the context of cryptocurrency exchanges. We also discussed how implementation of such security investments might be accomplished in real time through a horizontal-scaling mechanism that isolates traded volume into discrete segments. 
Our model and analyses do have certain limitations, which we intend to improve upon in future work.  First, our analyses only consider some of the day-to-day business decisions which guide the incentives of cryptocurrency exchanges.  Other major decisions relevant to cryptocurrency exchanges include a choice of exchange markets, as well as marketing and advertising strategies. More ambitiously, to truly capture the full risk dynamic that can result in the success or demise of an exchange, we would need to employ a longer-term analysis, that considers liquidity over time.  We believe that our overall framework is suitable for such an analysis, and that this is something worth pursuing in the future.
Finally, our analysis has focused on the decisions honest exchanges face when deciding how much effort to put towards security. We note that many of the highest-profile exchange failures have uncovered dishonest actions taken by exchange operators. For example, after Mt. Gox collapsed, evidence emerged that insiders manipulated the Bitcoin price by issuing unauthorized trades~\cite{Gandal18}. Bitconnect marketed itself as a trading platform and exchange serving a token valued at \$2.7 billion at its peak. However, it suddenly shut down in January 2018 and was subsequently revealed to be a Ponzi scheme~\cite{bitconnect}. The risk that the exchange itself could be malicious is very real, and one that future models could take into account.

To the best of our knowledge, this paper is the first to attempt modeling the incentives of cryptocurrency exchanges for security investments. Our development of a simplified model and our focus on only some aspects of day-to-day decisions made by exchanges 
 is justified in part by an acute observation that there does not exist a single model for cryptocurrency-exchange security investment strategies that is simultaneously applicable, analyzable, and efficiently communicable. As with progress on many other problems with real world limitations, however, efforts that help us better understand smaller pieces of complex interactions may usefully serve as good first steps.

\section*{Acknowledgments}
We thank the anonymous reviewers for their feedback. This work was supported in part by the US National Science Foundation under Award No. 1714291. The research activities of Benjamin Johnson and Jens Grossklags are supported by the German Institute for Trust and Safety on the Internet (DIVSI).

\balance
\bibliographystyle{splncs03}
\bibliography{references}

\end{document}